\documentclass[12pt]{article} 
\usepackage{epsf}

\usepackage{graphicx}
\newcommand{\beq}{\begin{equation}}
\newcommand{\eeq}{\end{equation}} 
\newcommand{\beqa}{\begin{eqnarray*}}
\newcommand{\eeqa}{\end{eqnarray*}}

\begin{document}
\begin{center}
{\bf Dissociation of Heavy Quarkonium in Quark-Gluon Plasma}
\end{center}
\vspace*{1cm}
\begin{center}
{\bf Sidi C. Benzahra$^{1}$, Benjamin F. Bayman$^{2}$}
\end{center}
\begin{center}
{\it $^{1}$ North Dakota State University}
\end{center}
\begin{center}
{\it Fargo, ND 58105, USA}
\end{center}
\begin{center}
{\it $^{2}$School of Physics and Astronomy, University of Minnesota}
\end{center}
\begin{center}
{\it Minneapolis, MN 55455, USA}
\end{center}

\vspace*{1cm}
\begin{abstract}
\noindent
In this work we calculate the lifetime of quarkonium 
moving with velocity $v$ through a quark gluon plasma at temperature $T$. We also investigate the stability of 
heavy mesons with respect to the effects of color charge screening. An explicit,
configuration-space potential is found for the screened interaction between the 
quarks constituting the meson.  We solve the Schr\"odinger equation for the
relative motion of the quarks in this non-spherical potential. In this way, we determine the range of $v,T$ values for which the meson is bound. When a bound state exists, we use the bound-state 
wavefunction as the initial state for the dissociation of the meson due to gluon absorption.  The meson 
lifetime is thus determined as a function of $v$ and $T$, and conclusions are drawn 
concerning the possibility of detection of the 
meson in a high-energy heavy-nucleus collision.

\end{abstract}

{\bf I. Introduction}\\

\hspace{0.5cm} Calculating the lifetime of a heavy meson in a medium of quarks and gluons is 
important for the understanding of the quark-gluon plasma.  If a heavy-nucleus collision were to produce 
fewer than the number of mesons expected, we would need to ask about the lifetime of the meson in 
comparison with the lifetime of the quark-gluon plasma.  If the lifetime of the quark-gluon plasma 
is greater than the lifetime of the meson in the quark-gluon plasma, one would think that there 
was probably a suppression of mesons due to the effect of screening or due to the effect of collision 
with gluons.  In fact, many claim that suppression of $J/\psi$ in 
heavy-ion collision could be a signature of quark-gluon plasma.\\

\hspace{0.5cm} There is a relationship between the inverse screening length, $m_{el}$; the number of flavors 
in the quark-gluon plasma, $N_{f}$; and the temperature of the quark-gluon plasma, $T$. If we 
increase the temperature of the quark-gluon plasma, the quarks and the gluons will become 
more active, and more effective at screening the interaction between a quark-antiquark pair. This relationship can be algebraically expressed [1] as 
follows:
\begin{equation}
{m_{el}^{2}} = {1 \over {3}} g^{2} (N + {{N_{f}} \over {2}}) T^{2} \, ,
\label{kapusta1}
\end{equation}
where $N=3$ from the color SU($N=3$) group, g is the dimensional coupling constant of the field 
strength, $F_{a}^{\mu \nu} = \partial^{\mu}A_{a}^{\nu}-\partial^{\nu}A_{a}^{\mu}-g 
f_{abc}A_{b}^{\mu}A_{c}^{\nu}$, and $N_{f}=3$ is the number 
of light flavors in the quark-gluon plasma, which are the up, the down, and the strange 
quarks. We will also use the temperature-dependent running coupling constant of QCD, which 
is given by [1, 2]
\begin{equation}
{{g^{2}} \over {4 \pi}} = {{12 \pi} \over {({11N -2N_{f}}) \rm \log \left({T^{2}}/\Lambda^{2} \right)}} \, .
\end{equation}
This equation explicitly displays asymptotic freedom: $g^2\rightarrow 0$ as $T \rightarrow \infty .$
We notice that there is no intrinsic coupling ``constant'' on the right hand side of this 
equation. The only free parameter of the theory is the 
QCD energy scale, $\Lambda$, whose numerical value is dependent on
the gauge and on the renormalization scheme chosen. If we choose the QCD energy scale to 
be $\Lambda=50$ MeV [1], $N_{f}=3$, and $N=3$ we get
\begin{equation}
m_{el}={{2 \pi T}\over{\sqrt{3 \rm \log(T/50)}}}\, .
\end{equation}

{\bf II. Screened Potential}\\

\hspace{0.5cm} Chu and Matsui [3] studied the effects of Debye screening of the interaction between a heavy quark-antiquark pair moving through a quark-gluon plasma. For our purposes, their results may be summarized in the form of an effective in $\vec{k}$ space quark-antiquark screened interaction:
\beq
V(k,\cos 
\theta_{{\bf k}})=\frac{k^2+A}{(k^2+A)^2+B^2}(1-\gamma^2(1-
\zeta^2))+\frac{k^2+C}{(k^2+C)^2+D^2}\gamma^2(1-\zeta^2),
\label{ii.4}
\eeq
where the quantities $\zeta,A,B,C,D$ are defined in terms of $\cos 
\theta_{{\bf k}}\equiv x$ by 
\beqa 
\zeta&\equiv&\frac{vx}{\sqrt{1-v^2(1-x^2)}} \\ 
A&\equiv&\frac{1}{2}\zeta^2+\frac{1}{4}(1-\zeta^2)\zeta \log 
\left(\frac{1+\zeta}{1-\zeta}\right) \\ 
B&\equiv&\frac{\pi}{4}(1-\zeta^2)\zeta \\ C&\equiv&(1-\zeta^2)(1-\frac{1}{2}\zeta \log \left(\frac{1+\zeta}{1-
\zeta}\right) \\
D&\equiv&\frac{\pi}{2}(1-\zeta^2)\zeta.
 \eeqa
$v$ is measured in units of $c$, and $k$ is measured in units of 
$m_{el}c/\hbar$, the inverse Compton wavelength associated with the 
screening mass, $m_{el}$. The resulting interaction potential is given in 
units of $\frac{4}{3} \alpha_s m_{el}c^2$, with $\alpha_s$ determined by QCD sum rules and lattice QCD calculations to be 0.232 [5]. The quantities $\zeta, A, B, C, D$ 
defined here are non-negative in the integration region. $V({\bf k})$ is invariant under rotation about ${\hat z}$, and under 
reflection across the ${\hat x}-{\hat y}$ plane.\\

\hspace{0.5cm} We obtain the quark-antiquark ${\bf r}$-space interaction potential, $U({\bf r})$, by Fourier transforming $V(\bf k)$:
\beq
U({\bf r})=\frac{1}{2 \pi^2}\int d^3k~e^{i {\bf k} \cdot {\bf 
r}}~V({\bf k})
\label{ii.2}
\eeq
In order to solve the quark-antiquark Schr\"odinger equation in ${\bf r}$ 
space, we need an expansion of the interaction potential in the form 
\beq 
U({\bf r})=\sum_\ell u_\ell(r)P_\ell(\cos \theta_{{\bf r}}) 
\label{ii.1} 
\eeq 
If we substitute the multipole expansion of the plane wave
\beqa
e^{i {\bf k} \cdot {\bf r}}&=&\sum_{ell}i^\ell (2 \ell+1) j_\ell 
(kr)P_\ell ({\hat k} \cdot{\hat r}) \\
~&=&4 \pi \sum_\ell i^\ell j_\ell(kr)\sum_\mu(Y^\ell_\mu({\hat 
k}))^*Y^\ell_\mu({\hat r})
\eeqa
into Equation (\ref{ii.2}), the axial symmetry of $V({\bf k})$ 
implies that only the $\mu=0$ terms will survive the $\phi_{\bf k}$ 
integration. Comparison with Equation (\ref{ii.1}) yields
\beq
u_\ell(r)=i^\ell\frac{2}{\pi}(2 \ell+1) \int_0^1 dx 
P_\ell(x)\int_0^\infty k^2 dk j_{\ell} (kr)V(k,x).
\label{ii.3}
\eeq
The integration variable $x$ in Equation (\ref{ii.3}) represents $\cos 
\theta_{{\bf k}}$. Because $V(k,x)$, given by Equation (\ref{ii.4}), is an even function of $x$, only even values of $\ell$ will occur in the multipole expansion.\\

\hspace{0.5cm} The $k$-integration in Equation (\ref{ii.3}) can be done exactly, using 
the theory of residues. We start with explicit expressions for the 
even-$\ell$ spherical Bessel functions:
\beq
j_\ell(kr)=\sum_{m=1,2,\ldots,\ell+1}\frac{(\ell+m-1)!}{(\ell-
m+1)!2^{m-1}(m-1)!}\times
\frac{a_m^\ell \sin(kr)+b_m^\ell \cos(kr)}{(kr)^m}
\label{ii.5}
\eeq
with
$$
a_m^\ell \equiv (-1)^{\frac{\ell+1-m}{2}},~~~~~~~~~~b_m^\ell \equiv 
0~~~~~~~~~~{\rm for~odd}~m
$$
$$
a_m^\ell \equiv 0,~~~~~~~~~~b_m^\ell \equiv (-1)^{\frac{\ell+s- m}{2}}~~~~~~~~~~{\rm for~even}~m $$ Individual terms of the sum in Equation (\ref{ii.5}) diverge as $kr 
\rightarrow 0$, but the entire sum converges as
$$
\lim_{kr \to 0}j_\ell(kr)=\frac{(kr)^\ell}{(2 \ell+1)!!}
$$

For the first term in Equation (\ref{ii.4}), we need the integral 
\begin{eqnarray} 
I_\ell &\equiv& \int_0^\infty k^2 dk 
\frac{k^2+A}{(k^2+A)^2+B^2}~j_{\ell}(kr)~~~~~({\rm even~}\ell) \label{ii.6} \\ &=&\frac{1}{2}\int_{-\infty}^\infty k^2 dk 
\frac{k^2+A}{(k^2+A)^2+B^2}j_{\ell}(kr) \nonumber \\ &=&I_\ell^{(+)}~+~I_\ell^{(-)},\nonumber
\end{eqnarray}
with
\begin{eqnarray*}
I_\ell^{(+)}&=&\frac{1}{4}\sum_{m=1}^{\ell+1}\frac{(\ell+m-1)!(-
ia_m^\ell+b_m^\ell)}{(\ell-m+1)!2^{m-1}(m-1)!r^m}\int_{-
\infty}^\infty\frac{e^{ikr}(k^2+A)}{[(k^2+A)^2+B^2]~k^{m-2}}dk\\
I_\ell^{(-)}&=&\frac{1}{4}\sum_{m=1}^{\ell+1}\frac{(\ell+m-
1)!(ia_m^\ell+b_m^\ell)}{(\ell-m+1)!2^{m-1}(m-1)!r^m}\int_{-
\infty}^\infty\frac{e^{-ikr}(k^2+A)}{[(k^2+A)^2+B^2]~k^{m-2}}dk.
\end{eqnarray*}
The contours used for the evaluation of these integrals are shown in 
Figures 1a and 1b.
In both cases, the integrand is vanishingly small on the infinite semi- circular parts of the contours. On the real k-axis, and on the 
semicircle around $k=0$, the sum of $I_\ell^{(+)}$ and $I_\ell^{(-)}$ 
gives us the convergent integrand we need for Equation (\ref{ii.6}). 
Thus we conclude that
\begin{eqnarray}
I_\ell&=&2 \pi i\times [{\rm 
(sum~of~residues~at~poles~within~the~contour~of~Figure~1a)}~-~\nonumber 
\\
&&{\rm (sum~of~residues~at~poles~within~the~contour~of~Figure~1b)]}.
\end{eqnarray}
\includegraphics[scale=0.5]{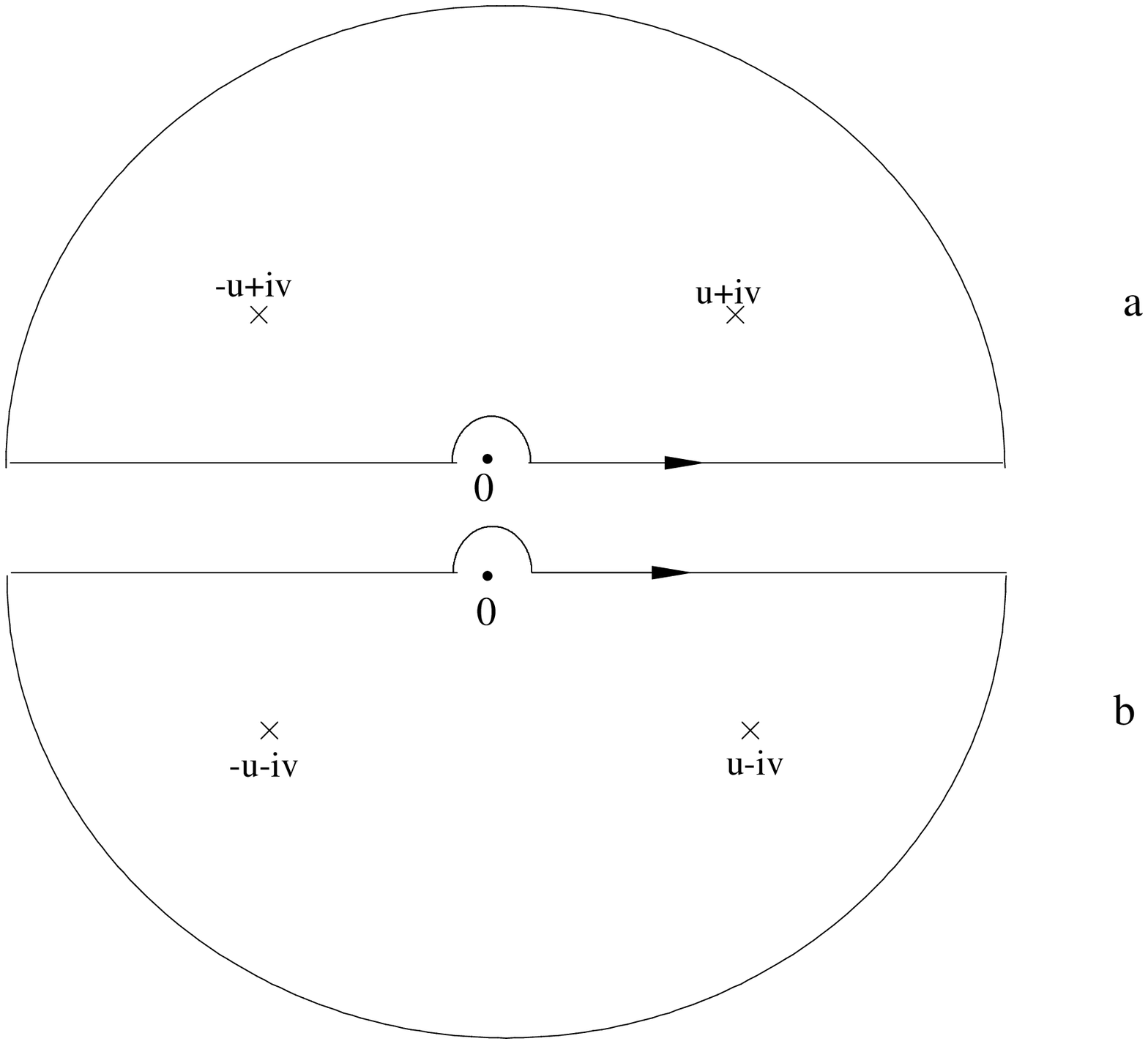}

\vspace{.5cm}
The poles within the contour of Figure 1a are at the zeroes of 
$(k^2+A)^2+B^2$ with positive imaginary parts, \textit{i.e} they are at 
$\pm u+iv$, where
$$
u \equiv \sqrt{\frac{\sqrt{A^2+B^2}-A}{2}},~~~~~~v \equiv 
\sqrt{\frac{\sqrt{A^2+B^2}+A}{2}}.
$$
The poles within the contour of Figure 1b are at $\pm u-iv$, and at 
$k=0$. The final result is
\begin{eqnarray}
I_\ell&=&\frac{\pi}{2}e^{-vr}\sum_{m=1}^{\ell+1}\frac{(\ell+m-
1)!\left(a_m^\ell\cos(ur-(m-1)\arctan \frac{v}{u})-b_m^\ell\sin(ur-(m- 1)\arctan \frac{v}{u})\right)}{(\ell-m+1)!2^{m-1}(m-
1)!(A^2+B^2)^{\frac{m-1}{4}}r^m} \nonumber \\
&+&X_\ell,
\end{eqnarray}
where
\begin{eqnarray*}
X_0~&\equiv &~0 \\
X_2~&\equiv &~\frac{\pi}{2} \frac{A}{A^2+B^2}\frac{3}{r^3} \\ X_4~&\equiv &~\frac{\pi}{2}\left[ \frac{A}{A^2+B^2} \frac{15}{2 
r^3}~+~\frac{B^2-A^2}{(A^2+B^2)^2} \frac{105}{r^5} \right]\\ X_6~&\equiv &~\frac{\pi}{2}\left[\frac{A}{A^2+B^2} \frac{105}{8 
r^3}~+~\frac{B^2-A^2}{(A^2+B^2)^2} \frac{945}{2r^5}~+~\frac{A(A^2- 3B^2)}{(A^2+B^2)^3}\frac{10395}{r^7} \right],{\cdots}~{\rm etc} 
\end{eqnarray*} 
Here $X_\ell$ represents the contribution to $I_\ell^{(-)}$ of the 
residue at the $k=0$ pole.\\

\hspace{0.5cm} The $k$ integral for the second term in Equation (\ref{ii.4}) is 
performed in the same way. Finally, it is necessary to do the $x$ 
integration in Equation (\ref{ii.3}).  This must be done numerically. 
However, since the range is finite $(0 \leq x \leq 1)$ and the 
integrand is smooth, the integrand can be accurately performed with 
relatively few points. The $x$-integrations in the results shown below 
used Simpson's rule, with an $x$-interval of 0.001.
An example of $u_\ell(r)$, for $v=0.5$ and $T=150 MeV$, is shown in Figure 2.\\

\includegraphics[scale=0.5]{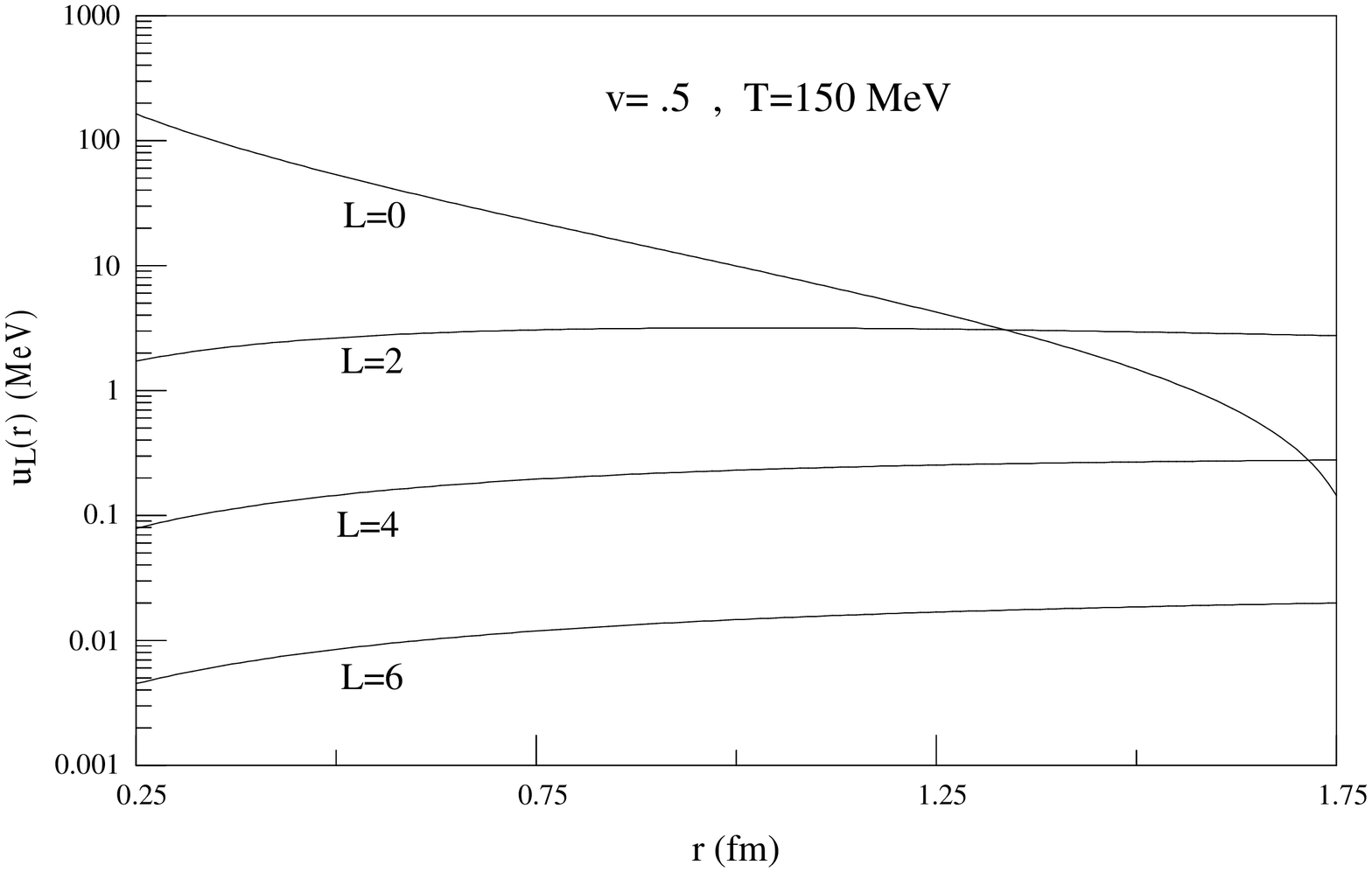}

{\bf III. Numerical Solutions of Schr\"odinger Equation}\\

\hspace{.5 cm} The binding energy of the quark and the antiquark can be 
calculated numerically by finding bound-state solutions of the Schr\"odinger equation.  In this case, the 
non-relativistic Schr\"odinger equation governing the relative motion of the 
two quarks is written in the form
\begin{equation}
[-\frac{\hbar^{2}}{2
\mu}\nabla^{2}~+~u_{0}(r)~+~u_{2}(r)P_{2}(cos{\theta})]~\psi(r,\theta,\phi)~=~E~
\psi(r,\theta,\phi) \, .
\label{bayman1}
\end{equation}
where $\mu=m_{Q}/2$ is the reduced $Q\bar{Q}$ mass.  We have included only $u_{0}(r)$ and  $u_{2}(r)$ in Equation (\ref{bayman1}) because, in the important $r$-region, $u_\ell(r)$ with $\ell>2$ are negligibly small (see Figure 2). Because the potential is axially-symmetric, the eigenfunctions will be 
characterized by a definite $m-$value. However, the spherical symmetry
of the potential is destroyed by the $u_{2}(r)P_{2}(cos{\theta})$ term, and
so the eigenfunctions will not be characterized by a unique value
of the total angular momentum. Thus a solution must be constructed as a linear
combination of total-angular-momentum eigenstates:  
\begin{equation}
\psi(r,\theta,\phi)~=~\sum_{\ell}\frac{\phi_{\ell}(r)}{r}~Y^{\ell}_{m}(\theta,
\phi) \, .
\end{equation}
If this expansion is substituted into Equation (\ref{bayman1}) the result will be a set of
coupled second-order differential equations for the radial
functions $\phi_{\ell}(r)$:
\begin{equation}
[~\frac{d^{2}}{dr^{2}}~-\frac{\ell (\ell +1)}{r^{2}}-\frac{2 \mu}
 {\hbar^{2}}~u_{0}(r)~-~k^{2}~]~\phi_{\ell}(r)~-~\frac{2 \mu}
 {\hbar^{2}}\sqrt{{4\pi}\over{5}}u_{2}(r)~\sum_{\ell '}~M_{\ell,\ell '}~\phi_{\ell '}(r)~=~0.
 \label{bayman3}
\end{equation}
In Equation (\ref{bayman3}) we have introduced the notation
\begin{eqnarray*}
E~&\equiv&~-\frac{\hbar^{2}}{2 \mu}~ k^{2} \\
M_{\ell,\ell'}~&\equiv&~\int \sin \theta d {\theta} d{\phi}~
(Y^{\ell}_{m}(\theta,\phi))^{*}~Y^{2}_{0}(\theta,\phi)~Y^{\ell
'}_{m}(\theta,\phi)~\\
~&=&~\sqrt{\frac{5(2 \ell'+1)}{4 \pi (2\ell +1)}}~(~2~\ell'~0~m~|~\ell~m~)~
(~2~\ell'~0~0~|~\ell~0~) \, .
\end{eqnarray*}
If we use explicit expressions for the vector-coupling coefficients, we obtain
\begin{eqnarray*}
M_{\ell,\ell'}&=&\frac{1}{2
\ell-1}\sqrt{\frac{45(\ell^{2}-m^{2})((\ell-1)^{2}-m^{2})}{16 \pi(2 \ell-3)(2
\ell+1)}}~~~~~{\rm if}~~\ell=\ell'+2 \\
~&=&\sqrt{\frac{5}{4 \pi}}~\frac{\ell(\ell+1)-3 m^{2}}{(2
\ell-1)(2\ell+3)}~~~~{\rm if}~~\ell=\ell' \\
 ~&=&\frac{1}{2
\ell'-1}\sqrt{\frac{45(\ell'^{2}-m^{2})((\ell'-1)^{2}-m^{2})}{16 \pi(2
\ell'-3)(2 \ell'+1)}}~~~{\rm if}~~\ell'=\ell+2
\end{eqnarray*}

\hspace{0.5cm} In order to describe bound states of the meson we must find 
normalizeable solutions of the coupled equations (\ref{bayman3}) which are regular at
$r=0$. This defines an eigenvalue condition for $k^{2}$.
A convenient numerical approach to this problem is suggested by the simple special case in which
$u_{2}(r)=0$. The coupling terms in Equation (\ref{bayman3}) vanish, leading to
the single equation
\begin{equation}
[~\frac{d^{2}}{dr^{2}}~-\frac{\ell (\ell +1)}{r^{2}}-\frac{2 \mu}
 {\hbar^{2}}~u_{0}(r)~-~k^{2}~]~\phi_{\ell}(r)~=~0.
 \label{bayman4}
\end{equation} 
We choose an arbitrary radius, $R$, and we guess a value for $k^{2}$. We then
find an {\it interior} solution $\phi_{\ell}^{i}(r)$ by numerically integrating
Equation (\ref{bayman4}) from $r=0$ to $r=R$, starting at $r=0$ with the behavior
\begin{equation}
\phi_{\ell}^{i}(r)~\buildrel r \rightarrow 0 \over \longrightarrow~r^{\ell +1}\, .
\end{equation}
Then we find an {\it exterior} solution $\phi_{\ell}^{e}(r)$ by numerically
integrating Equation (\ref{bayman4}) from a very large value of $r$ down to $r=R$, starting
at the large value of $r$ with the behavior
\begin{equation}
\phi_{\ell}^{e}(r)~\buildrel r \rightarrow \infty
\over \longrightarrow~h_{\ell}^{1}(ikr) \, .
\end{equation}
The eigenvalue condition on $k$ is that the relative normalizations of the
interior and exterior functions can be chosen so that there is no discontinuity
in value and slope at $r=R$. This requires that their logarithmic derivatives at
$r=R$ be equal, which we can expressed as the condition
\begin{equation}
\frac{\phi_{\ell}^{e}(R)}{\phi_{\ell}^{e~'}(R)}~\frac{\phi_{\ell}^{i~'}(R)}
{\phi_{\ell}^{i}(R)}~-~1~=0 \, .
\label{bayman7}
\end{equation}
$k^{2}$ in Equation (\ref{bayman4}) is varied until this condition is satisfied. The values
of $k^{2}$ so obtained are independent of the arbitrarily chosen $R$.  This can be seen by using (\ref{bayman4}) to show that
\begin{equation}
0=\phi_{\ell}^{e}{{d^{2}}\over{dr^{2}}}\phi_{\ell}^{i}-\phi_{\ell}^{i}{{d^{2}}\over{dr^{2}}}
\phi_{\ell}^{e}={{d}\over{dr}}\left [\phi_{\ell}^{e}{{d}\over{dr}}\phi_{\ell}^{i}-\phi_{\ell}^{i}{{d}\over{dr}} 
\phi_{\ell}^{e}\right ]\, ,
\end{equation}
so that $\phi_{\ell}^{e}{{d}\over{dr}}\phi_{\ell}^{i}-\phi_{\ell}^{i}{{d}\over{dr}} \phi_{\ell}^{e}$ is independent of r, 
and so if (\ref{bayman7}) is true at one value of r, it is true at all r.\\

\hspace{0.5cm} To generalize this procedure to the full set of coupled equations (\ref{bayman3}), we
define sets of interior and exterior functions by
\begin{eqnarray}
a_{\ell,\ell_{1}}(r)~\buildrel
r\rightarrow 0 \over \longrightarrow~\delta_{\ell,\ell_{1}}~r^{\ell +1}  \\
b_{\ell,\ell_{2}}(r)~\buildrel
r\rightarrow \infty \over
\longrightarrow~\delta_{\ell,\ell_{2}}~h_{\ell}^{1}(ikr) 
\end{eqnarray} 
We now
attempt to choose linear combinations of these interior and exterior functions
\begin{eqnarray}
\phi_{\ell}^{i}(r)=\sum_{\ell_{1}}~a_{\ell,\ell_{1}}(r)~\alpha_{\ell_{1}}\\
\phi_{\ell}^{e}(r)=\sum_{\ell_{2}}~b_{\ell,\ell_{2}}(r)~\beta_{\ell_{2}}
\end{eqnarray}
in order to achieve continuity of value and derivative at the
matching radius $r=R$. This requires that the coefficients $\alpha_{\ell_{1}}$
and $\beta_{\ell_{2}}$ satisfy
\begin{eqnarray}
\sum_{\ell_{1}}~a_{\ell,\ell_{1}}(R)~\alpha_{\ell_{1}}&=&\sum_{\ell_{2}}~
b_{\ell,\ell_{2}}(R)~\beta_{\ell_{2}} \label{bayman12}\\
\sum_{\ell_{1}}~a_{\ell,\ell_{1}}'(R)~\alpha_{\ell_{1}}&=&\sum_{\ell_{2}}~
b_{\ell,\ell_{2}}'(R)~\beta_{\ell_{2}}
\label{bayman13}
\end{eqnarray}
We can express this as a condition on the $\beta_{\ell}$ alone by using
Equation (\ref{bayman12}) to eliminate $\alpha_{\ell_{1}}$ from Equation (\ref{bayman13}):
\begin{equation}
\alpha_{\ell_{1}}=\sum_{\ell_{2}}~[~a^{-1}(R)~b(R)~]_{\ell_{1},\ell_{2}}~
\beta_{\ell_{2}}
\label{bayman14}
\end{equation}
\begin{equation}
\sum_{\ell_{2}}~[~b'^{-1}(R)~a'(R)~a^{-1}(R)~b(R)~]_{\ell_{1},\ell_{2}}~
\beta_{\ell_{2}}=\beta_{\ell_{1}}.
\label{bayman15}
\end{equation}
The necessary and sufficient condition for a non-trivial solution to Equation
(\ref{bayman15}) is
\begin{equation}
{\rm det}~[~b'^{-1}(R)~a'(R)~a^{-1}(R)~b(R)~-~{\bf 1}~]~=~0.
\label{baymanfix}
\end{equation}
This is the multi-channel generalization of Equation (\ref{bayman7}). The value of $k^{2}$
used in the numerical integration of the coupled equations (\ref{bayman3}) is varied until
Equation (\ref{baymanfix}) is satisfied to an acceptable accuracy. Then Equations (\ref{bayman14}) and
(\ref{bayman15}) determine the $\alpha_{\ell}$ and $\beta_{\ell}$ up to an overall
multiplicative factor, which can be obtained from the normalization
condition
\begin{equation}
\sum_{\ell}~\int_{0}^{R}dr~[~\sum_{\ell_{1}}a_{\ell,\ell_{1}}(r)
\alpha_{\ell_{1}}~]^{2}~+~  
\sum_{\ell}~\int_{R}^{\infty}dr~[~\sum_{\ell_{2}}b_{\ell,\ell_{2}}(r)
\beta_{\ell_{2}}~]^{2}~=~1.
\end{equation}
As in the one-channel case, the consistency of the procedure guarantees that the
calculated energy eigenvalues and eigenfunctions are independent of the choice
of the matching radius.\\

\hspace{.5 cm}The numerical integration of the coupled differential equations was performed
using the 4th-order Runge-Kutta method with a step-length of 0.01 fm, starting
at a minimum radius of 0.001 fm [4]. Although the $\ell$ sum in Equation (\ref{bayman3}) has, in
principle, no upper limit, the sum must be truncated in order to make the
calculation finite. In the results to be shown below, the upper limit on $\ell$
was taken to be 8. In fact, the strength of the coupling potential $u_{2}(r)$ is
small enough so that there was very little mixing of $\ell \ne 0$ components into
predominantly $\ell=0$ eigenfunctions.\\

\hspace{0.5cm} The binding energies obtained in this way are plotted in Figure 3, as functions of $v$ and $T$. Here we have used a heavy-quark mass of $m_{Q}=4.3 GeV/c^2$. It is seen that our calculations indicate that bound mesons cannot exist at temperatures greater than about $T=275$ MeV. Furthermore, at any given temperature, the binding energy is a decreasing function of $v$. \\

\includegraphics[scale=0.5]{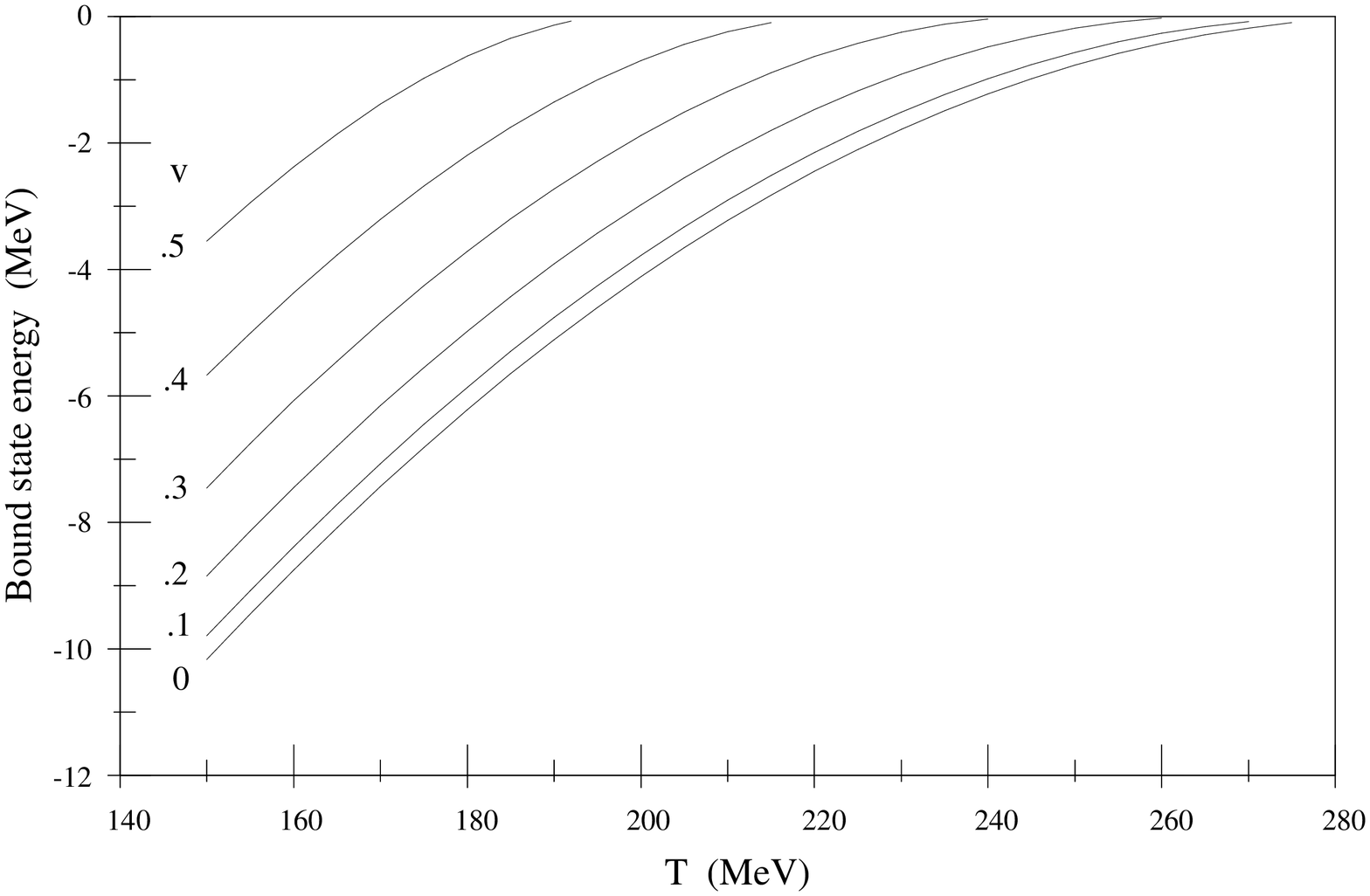}\\

{\bf IV. Dissociation Time of a Heavy Meson Due To Collision With Thermal 
Gluons}\\

\hspace{0.5cm}Let us introduce the calculation of the dissociation time by asking: What is the rate of photo-ionization if we place a hydrogen atom in a cavity of thermalized photons at 
temperature $T$?  The Planck distribution [6] states that the flux of photons in the 
interval $dk$ around $k$ is

\begin{equation}
{c{k^2 dk}\over {\pi^2 [e^{{\hbar k c}/ {T}}-1]}}\, .
\end{equation} 
Let $\sigma(k)$ be the cross-section for photo-emission when the incident photon has momentum $k$.  The rate for this process is then
\begin{equation}
R=\int_{0}^{\infty}{c{k^2 \sigma(k)} \over {\pi^2 [e^{{\hbar k c}/ {T}}-1]}}dk\, .
\end{equation} 
According to Gasiorowicz, Quantum Physics, Chapter 24 [7]
\begin{equation}
\sigma(k)=   {{V p_{e} e^{2}}\over{2 \pi \hbar^{3} m c^2 k}} \int d \Omega \left | \int d^{3}r \psi_{f}^{*}(\vec{r}) 
 \hat{\epsilon} \cdot \vec{p} e^{ikz}
\psi_{i}(\vec{r}) \right |^2 \, .
\end{equation}
$\psi_{i}(\vec{r})$ is the normalized wave function of the initial bound electron state, $p_{e}$ is the asymptotic momentum 
of the outgoing electron [8], and m is its mass.
\begin{equation}
\psi_{f}(\bar{r})~\buildrel r \rightarrow \infty
\over \longrightarrow~{1\over \sqrt{V}} \left [ e^{i \vec{k}_{e} \cdot \bar{r}} + f(\theta^{\prime}, \phi^{\prime}) {e^{-ik_{e}r } 
\over{r}}\right ] \, .
\end{equation}
\[{\vec{k}}_{e}={{{\vec{p}}_{e}}\over{\hbar}}=k_{e}[{\rm sin}{\theta^{\prime}}({\rm cos}\phi^{\prime} \hat{x} +{\rm
sin} \phi^{\prime} \hat {y})+{\rm cos} \theta^{\prime} \hat {z}]. \]
The $d \Omega$ integration is over $\theta^{\prime}$, $\phi^{\prime}$, the asymptotic direction of the outgoing electron.\\

\hspace{0.5cm} Let us now consider the transition from QED to QCD.  In QED the electron-photon vertex [9] is associated with matrix 
element of
\begin{equation}
i g_{e} \gamma^{\mu}=i \sqrt{{4 \pi e^{2}}\over{\hbar c}} \gamma^{\mu} \, .
\end{equation}
In QCD the quark-gluon vertex [9] is associated with matrix element of
\begin{equation}
-i {{g_{s}}\over{2}} \lambda^{\alpha} \gamma^{\mu}=-i \sqrt{{4 \pi \alpha_{s}}\over{2}} \lambda^{\alpha} \gamma^{\mu} \, .
\end{equation}
The $\lambda^{\alpha}$ is a generator of color SU(3). ($\alpha$ =1,2,...,8).  We are going from a meson 
(color singlet) to a $Q\bar{Q}$ octet.
\[<[{\rm octet}]_{\alpha}|\lambda^{\alpha}|[{\rm singlet}]>=\sqrt{{8}\over{3}}.\]
Thus 
\[ i \sqrt{{4 \pi e^{2}}\over{\hbar c}}\longleftrightarrow -i \sqrt{{4 \pi \alpha_{s}}\over{2}}
\sqrt{{8}\over{3}},\] 
which gives us
\[{e^{2}\over{\hbar c}} \longleftrightarrow {{2}\over{3}}\alpha_{s}\]
The photon flux density is multiplied by 8 to get the total gluon density 
(8 types of gluons, 8 generators of SU(3)). 
Finally,
\begin{equation}
R=\int_{0}^{\infty}{8 c{k^2 \sigma(k)} \over {\pi^2 [e^{{\hbar k c}/ {T}}-1]}}dk\, ,
\label{rate1}
\end{equation} 
and 
\begin{equation}
\sigma(k)=   {{V p_{Q} {{2}\over{3}} \alpha_{s}}\over{2 \pi \hbar^{2} \mu c k}} \int d \Omega \left 
| \int d^{3}r \psi_{f}^{*}(\vec{r}) \hat{\epsilon} \cdot \vec{p} e^{ikz} \psi_{i}(\vec{r}) \right |^2 \, ,
\end{equation}
where $\mu$ is the reduced $Q\bar{Q}$ mass.\\

\hspace{0.5cm} In his evaluation of $\sigma(k)$, B. Muller [10] replaced $e^{ikz}$ by 1 ( long-wave-length approximation) and $\psi_{f}(\vec{r})$ by $e^{i
{\vec{k}}_{Q}\cdot \vec{r}}$ (no distortion of the outgoing wave) and neglected the connection between $k_{Q}$ and $k$.  We will make
none of these approximations. \\

\hspace{0.5cm} Now let us expand the final outgoing quark wavefunction
\begin{equation}
\psi_{f}(\vec{r}) = {{4 \pi} \over {\sqrt{V}}}\sum_{\ell} i^{\ell} e^{-i\delta_{\ell}} \sum_{m}{Y_{m}^{\ell}}^{*}
(\theta^{\prime}, \phi^{\prime}) Y_{m}^{l}(\theta,\phi) {{W_{l}(r)} \over r}\, .
\label{fwavefunction1}
\end{equation}
\begin{eqnarray}
\hat{\epsilon}\cdot \vec{p} e^{ikz}\psi_{i}(\vec{r})&=&  {{\hbar} \over {i}} \hat{\epsilon} \cdot \vec{\nabla}e^{ikz}\psi_{i}(r)\\ 
       				                &=&{{\hbar} \over{i}} e^{ikz}\hat{\epsilon} \cdot \vec{\nabla}\psi_{i}(r)
						\nonumber \\
						&=&{{\hbar} \over{i}} e^{ikz}\hat{\epsilon}
						\cdot \hat{r}{\psi_{i}}^{\prime}(r) \nonumber 						
\end{eqnarray}
The last three equalities are satisfied since $\hat{\epsilon} \cdot \hat{z}=0$ (gluon has transverse
polarization) and since $\psi_{i} (\vec{r})\approx \psi_{i}(r)$ (nearly spherically symmetric ground state). The $W_{l}(r)$ are calculated using the methods of Section III. \\

\hspace{0.5cm} Let 
\begin{equation}
I({\vec{k}}_{Q}) = \int d^{3}r \psi_{f}^{*}(\vec{r}) \hat{\epsilon}\cdot \vec{p} e^{ikz} \psi_{i}(\vec{r}) \, ,
\end{equation}
which is the same as
\begin{equation} 
I({\vec{k}}_{Q})= {{\hbar}\over{i}} {{4 \pi}\over{\sqrt{V}}} \sum_{l,m} i^{-\ell} e^{i \delta_{\ell}} Y_{m}^{l}
(\theta^{\prime}, \phi^{\prime}) \int d^{3}r {Y_{m}^{l}}^{*}(\theta, \phi){{W_{l}(r)}\over{r}} \hat{\epsilon}\cdot \hat{r} e^{ikz} 
\psi_{i}^{\prime}(r) \, ,
\label{baymanI1}
\end{equation}
where 
\[ \hat{\epsilon}\cdot \hat{r}=\epsilon_{x} {\rm sin}\theta {\rm cos}\phi + \epsilon_{y} {\rm sin} \theta {\rm sin} \phi=
\sqrt{{2 \pi}\over{3}}[-(\epsilon_{x}-i\epsilon_{y})Y_{1}^{1}(\theta, \phi)+ (\epsilon_{x}+i\epsilon_{y})
Y_{-1}^{1}(\theta, \phi)] \, .\]
Now we use the relation
\begin{equation}
Y_{\pm 1}^{1}(\theta, \phi)e^{ikz}=\sqrt{3\over2}\sum_{\ell}i^{\ell-1}\sqrt{\ell (2\ell+1)(\ell+1)}
{{j_{\ell}(kr)}\over{kr}} 
Y_{\pm 1}^{\ell}(\theta, \phi) \, .
\label{baymanY1}
\end{equation}
Substituting Equation (\ref{baymanY1}) in Equation 
(\ref{baymanI1}), we get
\begin{eqnarray}
I({\vec{k}}_{Q})&=& {{4 \pi^{3/2} \hbar}\over{\sqrt{V}k}} \sum_{\ell=1}^{Lmax} e^{i \delta_{\ell}}\sqrt{\ell (2\ell+1)(\ell+1)} \times \\
                & &  [(\epsilon_{x}-i\epsilon_{y}) Y_{1}^{1}(\theta^{\prime}, \phi^{\prime})- (\epsilon_{x}+i\epsilon_{y}) 
		Y_{-1}^{1}
(\theta^{\prime}, \phi^{\prime})]\int_{0}^{\infty} dr j_{\ell}(kr)W_{\ell}(r)\psi_{i}^{\prime}(r)\, .
\nonumber
\end{eqnarray}
When we integrate over $\theta^{\prime}$ and $\phi^{\prime}$ (the $d\Omega$ integration of Equation (37)), the orthonormality of the spherical harmonics converts the coherent sum over $\ell$ in $I({\vec{k}}_{Q})$ into an incoherent sum over $\ell$ in $\sigma(k)$
 
\begin{equation}
\sigma(k)={{32}\over{3}} \pi^{2} \alpha_{s} {{\hbar c}\over{m_{Q} c^{2}}} {{k_{Q}\over{k^{3}}}} \sum_{\ell=1}^{Lmax} 
{\ell (2\ell+1)(\ell+1)} \left |\int_{0}^{\infty} dr j_{\ell}(kr)W_{\ell}(r)\psi_{i}^{\prime}(r)\right |^{2} \, .
\label{sigma1}
\end{equation}
Equation (\ref{sigma1}) is inserted into Equation (\ref{rate1}), and the 
integration over $k$ is carried out numerically. Figure 4 shows a plot
of calculated lifetime $\tau=(1/R)$ as a function of $T$ and $v$.\\

\includegraphics[scale=0.5]{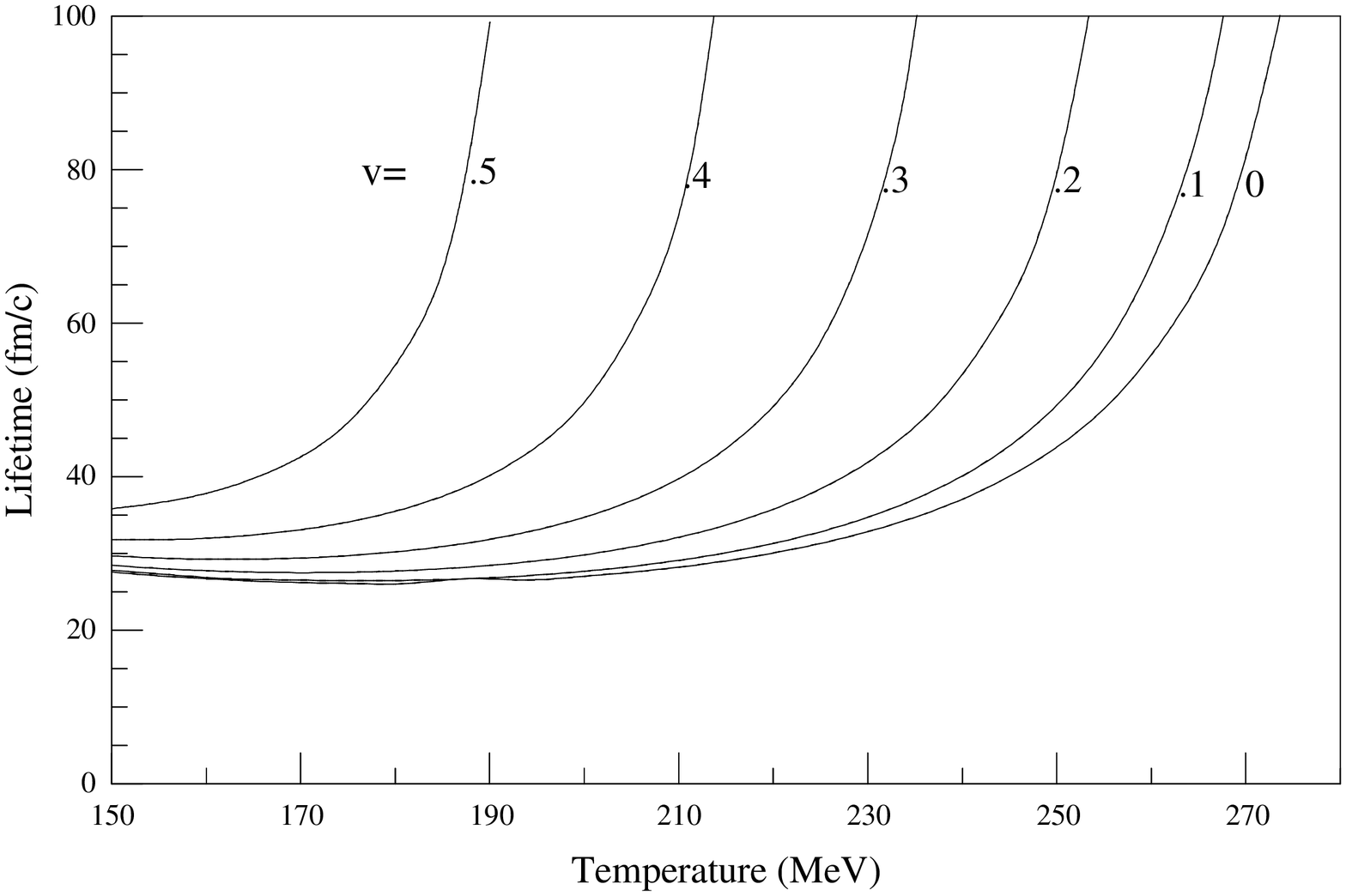}

\hspace{0.5 cm} The most significant feature of Figure 4 is the prediction of a \textit{minimum} lifetime. When $T$ decreases below about $T\sim170$ MeV, the disintegration decreases because of falling gluon flux. As $T$ increases above $T\sim170$ MeV, the disintegration rate decreases because of falling $\sigma (k)$. The maximum disintegration rate at $T\sim170$ MeV corresponds to a lifetime of $\sim 27.5$ fm/c.\\

\hspace{0.5cm} The decrease of $\sigma (k)$, in the important $k$ region, with increasing $T$, is due to requirements of conservation of energy and momentum in a process in which a gluon is absorbed and a quark is emitted. For larger $k$, it is increasingly difficult to find, in the meson inital state, quark momenta that are large enough to satisfy both energy and momentum conservation. An analagous phonomenon occurs in the photoelectric effect, and results in photoelectric cross-sections that decrease with increasing photon momentum, once a threshold has been passed. In our situation, this effect is exacerbated by the decrease of meson binding energy with increasing temperature (Figure 3). This results in a meson wave function with greater spatial extent, and thus a lower probability for the occurrence of the necessary high momentum components in the quark-antiquark wave function.\\ 

\hspace{0.5cm} No lifetime minimum with respect to $T$ was predicted 
by Muller [10] since he did not take into account both momentum and energy 
conservation.  The expected lifetime of the quark gluon plasma is 10-20 fm/c, which 
is appreciably shorter than our minimum predicted meson-disintegration lifetime of 27.5 fm/c.  Thus our 
conclusion is that the plasma will cease to exist before a significant 
fraction of the mesons will have been disintegrated by gluon impact.  Therefore, this process 
will not have an appreciable effect on the number of mesons observed.\\

{\bf V. Relation of Results to Experiment}\\

\hspace{.5cm} Recent experiments [11] at the CERN SPS have shown a large suppression of $J/\psi$ production in central Pb+Pb collisions. 
Following the original idea of Matsui and Satz [12] that $J/\psi$ would be dissociated in a quark-gluon plasma due to
color screening, the observed $J/\psi$ suppression has been suggested as an evidence for the formation of the quark-gluon 
plasma in these collisions [13-15]. \\

\hspace{.5cm} Since the $\Upsilon$ meson states in a quark-gluon plasma are also sensitive to 
color screening [12, 16, 17], the study of the $\Upsilon$ meson suppression in high energy heavy ion collisions can be 
used as a signature for the quark-gluon plasma as well.  Because the binding energy of $\Upsilon$ is larger than that 
of $J/\psi$, the critical energy density at which an upsilon meson is dissociated in the quark-gluon plasma is also 
higher [18]. One therefore expects to see the effects of the quark-gluon plasma on the production of the $\Upsilon$ meson 
only in ultra-relativistic heavy ion collisions such as at the RHIC and the LHC.  As in the case of $J/\psi$, one needs to 
understand the effects of the $\Upsilon$ meson absorption in hadronic matter in order to use its suppression as a signal 
for the quark-gluon plasma in heavy ion collisions.\\

\hspace{.5cm} It is shown in Figure 3 that a bound $\Upsilon$ meson cannot exist above a temperature of about 275 MeV, no matter what velocity it has relative to the plasma. Thus observation of $\Upsilon$ mesons implies that the $\geq 275$ MeV plasma never existed, or that the mesons were created in a plasma-free region. Bound $\Upsilon$ mesons can exist at lower temperatures, even at speeds relative to the plasma that are an appreciable fraction of $c$. Thus the complete absence of $\Upsilon$ mesons in a situation in which they should have been created is an indication of the existence of a $T>275$ MeV plasma.\\ 

\hspace{.5cm} In these considerations, it is not necessary to be concerned about dissociation of an $\Upsilon$ meson due to absorption of a gluon, in a process analagous to the photoelectric effect or the photo-disintegration of the deutron. Our calculations show that the lifetime of this process is longer than the expected lifetime of the plasma, so once a bound $\Upsilon$ is formed, it has a high probability of surviving the collision.     

\newpage

{\bf References}\\

[1] J.I. Kapusta, Finite Temperature Field Theory (Cambridge University Press, Cambridge, 
1989).\newline
[2] J.I. Kapusta, Phys. Rev. D {\bf 46}, Number 10, 4749 (1992).\newline
[3] M.C.Chu and T. Matsui, Phys. Rev. D {\bf 39}, 1892 (1989).\newline
[4] Numerical Recipes in Fortran 97, William H. Press, Saul A. Teukolsky, 
William T. Vetterling and Brian P. Flannery, Second Edition, Chapter 16,
Cambridge University Press, 1992 .\newline
[5] Matthias Jamin and Antonio Pich, Nucl. Phys. B $\bf 507$ (1997) 334-352. \newline
[6] F. Reif, Fundamentals of Statistical and Thermal Physics (McGraw-Hill, 
Inc. Singapore, 1965). \newline
[7] S. Gasiorowicz, Quantum Mechanics Second Edition (John Wiley and Sons, Inc. 1996). \newline
[8] G. Breit and H. A. Bette, Phys. Rev. $\bf 93$ (1954) 888 .\newline
[9] F. Halzen and A. Martin, Quarks and Leptons (John Wiley and Sons, Inc. 1984). \newline
[10] B. Muller, preprint nuc-th/9806023, v2 (unpublished).\newline
[11] M. Gonin {\it et al.}, the NA50 Collaboration, Nuc. Phys. A $\bf 610$ (1996) 404c; M.C. Abreu {\it et al.}, the NA50
Collaboration, Phys. Lett. B $\bf 450$ (1999) 456. \newline
[12] T. Matsui and H. Satz, Phys. Lett. B $\bf 178$ (1986) 416. \newline
[13] J. -P. Blaizot and J.-Y. Ollitrault, Phys. Rev. Lett. $\bf 77$ (1996) 1703. \newline
[14] C. -Y Wong, Nucl. Phys. A $\bf 630$ (1998) 487. \newline
[15] D. Kharzeev, M. Nardi and H. Satz, Phys. Lett. B $\bf 405$ (1997) 14; D. Kharzeev, 
C. Lourenco, M. Nardi and H. Satz, Z.\newline
[16] S. C. Benzahra, Phys. Rev. C $\bf 61$ 064906 (2000).\newline
[17] For recent reviews, see, e.g., R. Vogt, Phys. Rept. $\bf 310$ (1999) 197; H. Satz, Rept. Prog. Phys. $\bf 63$ (2000) 1511.\newline
[18] F. Karsch, M.T. Mehr and H. Satz, Z. Phys. C $\bf 37$ (1988) 617. \newline

\newpage

{\bf Figure Captions}\\

Figure 1.  Contours for the evaluation of $I^{(\pm)}$.\\

Figure 2.  $u_\ell(r)$ for $T=150 MeV$ and $v=0.5$\\

Figure 3.  Calculated meson binding energies as a function of $T$, for various values of $v$.\\

Figure 4.  Calculated meson dissociation lifetimes as a function of $T$, for various values of $v$.\\

\vspace{1cm}
{\bf Acknowledgements}\\
The authors would like to thank Joseph I. Kapusta for his very useful help.\\  

Part of this work is supported by the U.S. Department of Energy under contract DE-AC03-76SF00098,
and grant DE-FG02-87ER40328.\\

\end{document}